%% file: sample.tex
\documentclass[manuscript,screen,acmsmall,nonacm]{acmart}
\input{dlabmacros}
\usepackage{subcaption}
\usepackage[utf8]{inputenc}

\newcommand{\numDeplatformedinfluencers}{101\xspace}
\newcommand{\numDeplatformedEvents}{165\xspace}

\AtBeginDocument{%
  \providecommand\BibTeX{{%
    \normalfont B\kern-0.5em{\scshape i\kern-0.25em b}\kern-0.8em\TeX}}}

\setcopyright{acmcopyright}
\copyrightyear{2018}
\acmYear{2018}
\acmDOI{XXXXXXX.XXXXXXX}

\usepackage{tikz}
\usetikzlibrary{positioning, calc, shapes.geometric, shapes, shapes.multipart, arrows.meta, arrows, decorations.markings, external, trees}
\tikzset{
node distance=0.5cm, 
}
\tikzstyle{Arrow} = [
	thick, 
	decoration={
		markings,
		mark=at position 1 with {
			\arrow[thick]{latex}
			}
		}, 
	shorten >= 0.5pt, preaction = {decorate},
	]

\begin{document}

\title{Deplatforming Norm-Violating Influencers on Social Media Reduces Overall Online Attention Toward Them}
\author{Manoel Horta Ribeiro}
\affiliation{%
  \institution{EPFL}
  \country{Switzerland}}
\email{manoel.hortaribeiro@epfl.ch}

\author{Shagun Jhaver}
\affiliation{%
  \institution{Rutgers University}
  \country{USA}}
\email{sj917@comminfo.rutgers.edu}

\author{Jordi Cluet i Martinell}
\affiliation{%
  \institution{EPFL}
  \country{Switzerland}}
\email{jordi.cluetimartinell@epfl.ch}

\author{Marie Reignier-Tayar}
\affiliation{%
  \institution{EPFL}
  \country{Switzerland}}
\email{marie.reignier@epfl.ch}

\author{Robert West}
\email{robert.west@epfl.ch}
\affiliation{%
  \institution{EPFL}
  \country{Switzerland}}
\email{robert.west@epfl.ch}

\renewcommand{\shortauthors}{AnonAnon.}

\begin{abstract}
From politicians to podcast hosts, online platforms have systematically banned (``deplatformed'') influential users for breaking platform guidelines. 
Previous inquiries on the effectiveness of this intervention are inconclusive because
1) they consider only few deplatforming events;
2) they consider only overt engagement traces (e.g., likes and posts) but not passive engagement (e.g., views);
3) they do not consider all the potential places users impacted by the deplatforming event might migrate to.
We address these limitations in a longitudinal, quasi-experimental study of \numDeplatformedEvents deplatforming events targeted at \numDeplatformedinfluencers influencers.
We collect deplatforming events from Reddit posts and then manually curate the data, ensuring the correctness of a large dataset of deplatforming events.
Then, we link these events to Google Trends and Wikipedia page views, platform-agnostic measures of online attention that capture the general public's interest in specific influencers.
Through a difference-in-differences approach, we find that deplatforming reduces online attention toward influencers. 
After 12 months, we estimate that online attention toward deplatformed influencers is reduced by
$-$63\% (95\% CI [$-$75\%,$-$46\%]) on Google and by 
$-$43\% (95\% CI [$-$57\%,$-$24\%]) on Wikipedia.
Further, as we study over a hundred deplatforming events, we can analyze in which cases deplatforming is more or less impactful, revealing nuances about the intervention.
Notably, we find that both permanent and temporary deplatforming reduce online attention toward influencers;
Overall, this work contributes to the ongoing effort to map the effectiveness of content moderation interventions, driving platform governance away from speculation.

\end{abstract}

\ccsdesc[500]{Information systems~Social networking sites}
\ccsdesc[500]{Human-centered computing~Empirical studies in collaborative and social computing}
\ccsdesc[500]{Information systems~Social networks}

\keywords{online communities; fringe online communities; content moderation; online radicalization; deplatforming; social networks}

\received{20 February 2007}
\received[revised]{12 March 2009}
\received[accepted]{5 June 2009}

\maketitle
\newpage

\begin{figure}
    \centering
    \includegraphics[scale=0.9]{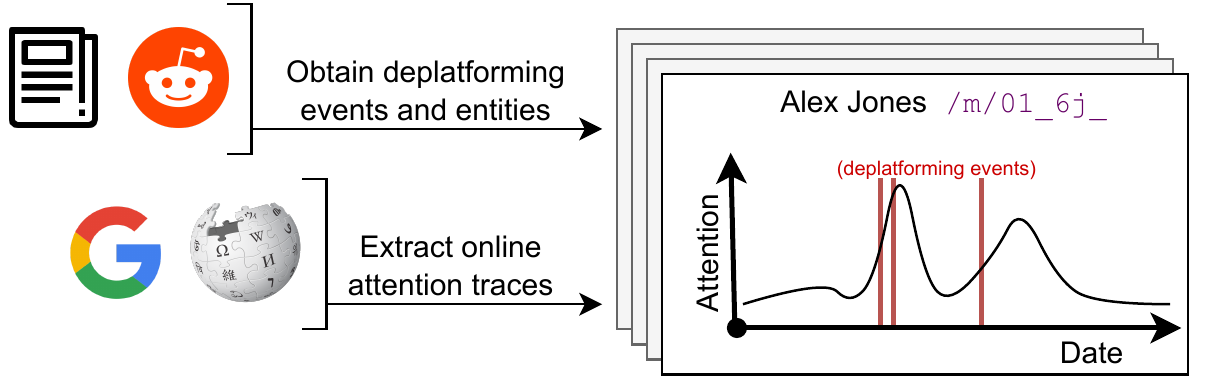}
    \caption{\textbf{Our approach to studying deplatforming.} 
    We obtain deplatforming events and entities from news shared on Reddit, e.g., \textit{Twitter bans Alex Jones and InfoWars; cites abusive behavior}~\cite{schneider2018twitter}. 
    Then, we extract the Google Knowledge Graph identifier of these entities (e.g., Alex Jones corresponds to \textit{/m/01\_6j\_}) and obtain digital attention traces from Wikipedia (page views) and Google Trends (search interest). Last, we analyze these time series with descriptive and quasi-experimental methods.}
    \label{fig:my_label}
\end{figure}

\section{Introduction}

To \emph{deplatform} is ``to remove and ban (a registered user) from a mass communication medium (such as a social networking or blogging website)''~\cite{meriamwebster_deplatform}.
The term has gained notoriety in recent years as a roster of provocateurs, extremists, and conspiracy theorists (often from the far right) were banned from online platforms like YouTube~\cite{yt_bans}, Twitter~\cite{tw_bans}, and Facebook~\cite{fb_bans}.
The practice has also become a contentious political issue.
On the one hand, deplatforming advocates suggest that it could minimize the consequences of users' exposure to inappropriate content on online platforms, often pointing at the negative role that platforms have played in the rise of science denial~\cite{covid_infodemic}, political extremism~\cite{radicalization}, \etc
On the other hand, opponents worry about the lack of accountability and transparency in platforms' deplatforming decisions and raise concern that this practice may silence dissenting but valid viewpoints~\cite{luo2022should}.

Importantly, deplatformed individuals are reactive and, following sanctions, migrate to alternative platforms~\cite{horta2021platform,rogers2020deplatforming}.
Websites like Gab, Rumble, and Truth Social have capitalized on the banning of prominent influencers (as well as the people outraged by it), advertising themselves as `censorship-free' and welcoming of extreme viewpoints~\cite{freelon2020false}.
Zuckermann and Niccoluci~\cite{zuckerman2021deplatforming} argue these migrations help solidify the infrastructure of an ``alt-tech'' information ecosystem, where the visibility of extreme content is reduced, but extreme viewpoints resonate louder.

Empirical evidence on whether deplatforming works is inconclusive. 
Previous work has explored this question in various settings: 
from studying how fans of controversial figures like Alex Jones behaved on Twitter after their bans~\cite{jhaver2021evaluating} to examining how entire communities re-organized themselves after being banned from Reddit~\cite{horta2021platform}. 
This literature (described in \Secref{sec:rw}) has found that deplatforming reduces the number of posts~\cite{horta2021platform}, tweets~\cite{jhaver2021evaluating}, or videos~\cite{rauchfleisch2021deplatforming} related to the deplatformed individuals or communities, albeit at the cost of an increase in the harmfulness of the remaining content~\cite{horta2021platform,ali2021understanding,mitts2021banned} often hosted on alt-tech platforms or standalone websites.

However, three important limitations of the existing literature threaten the validity and generalizability of its findings. 
Previous work typically
(1)~considers only a handful of deplatforming events (often a single one)~\cite{horta2021platform, jhaver2021evaluating, mitts2021banned, trujillo2022make, trujillo_one_2023, russo2023spillover, horta2023deplatforming, thomas_disrupting_2023};
(2)~considers only overt engagement traces like posts, but not passive engagement traces like impressions~\cite{russo2023spillover, horta2021platform, jhaver2021evaluating, mitts2021banned, trujillo2022make, trujillo_one_2023};
(3)~does not consider all the potential places users might migrate to~\cite{rauchfleisch2021deplatforming, russo2023spillover, urman2022they, horta2021platform, jhaver2021evaluating, mitts2021banned, trujillo2022make, trujillo_one_2023, thomas_disrupting_2023}.
Given these limitations, work capturing the heterogeneity of the effect (\ie, how the intervention works against different targets) and considering the impact of deplatforming across the Web is needed to provide adequate policy guidance on the matter~\cite{horta2023deplatforming}.

\xhdr{Present work} 
In this paper, we present a longitudinal analysis of \numDeplatformedEvents deplatforming events concerning  \numDeplatformedinfluencers influential users on social media, carefully sourced and annotated at both the entity level (\ie, who was the influencer---a politician or a media personality?) and at the event-level (\eg, why was the influencer deplatformed---for harassment or for spreading false information?).
We consider online attention toward these influencers, considering two easily accessible, freely available data sources: Wikipedia pageviews and Google search interest~\cite{gtab}.

We obtain our data from a comprehensive data collection pipeline. 
First, we use a semi-supervised approach to collect deplatforming events from Reddit posts. We match each deplatformed entity with a Google knowledge graph identifier that allows us to obtain online attention data for these entities on Wikipedia and Google Trends.
Second, we ensure the completeness and correctness of the deplatforming events for the entities considered by retrieving news sources associated with each entity and individually attributing a news piece confirming each platforming event.
Finally, we manually label each deplatforming event, specifying, \eg, the reason behind the intervention.
We make our dataset, the largest collection of deplatforming events (and associated attention traces) to date, and the code to reproduce our analyses publicly available.%
\footnote{\url{https://anonymous.4open.science/r/dep_influencers-D915/}}

We use a stacked difference-in-differences regression to disentangle the causal effects of deplatforming from the reason behind the intervention, following Cengiz et al.~\cite{cengiz2019effect}. 
This approach allows us to compare treated units (deplatformed influencers) with `yet-to-be-treated' control units (influencers who will be deplatformed in the future) and to identify the causal effect of deplatforming under the parallel-trends assumption, 
\ie, that in the absence of treatment, the differences between the two groups would remain constant.
Further, we address the limitations of previous research as we study passive engagement traces that are agnostic to specific social media platforms and consider orders of magnitude more deplatforming events than past research.
This has two implications.
First, we can better answer the question: does deplatforming reduce online attention toward influencers? 
Second, we can study whether different individuals are affected differently; \eg, do influencers banned for different reasons respond differently to being deplatformed?

\xhdr{Summary of results} 
We find that deplatforming reduces online attention toward influencers. 
Specifically, after 12 months, we estimate that online attention toward deplatformed influencers is reduced by
$-$63\% (95\% CI [$-$75\%,$-$46\%]) on Google and by 
$-$43\% (95\% CI [$-$57\%,$-$24\%]) on Wikipedia.
We also find that the effect of deplatforming varies according to the characteristics of the deplatformed entity.
For example, sorting entities into two groups according to the attention they received 12 months before deplatforming (top 1/3rd of influencers with highest attention 
\vs bottom 2/3rds),
we find that the effect of deplatforming for high-attention entities was roughly 60\% lower relative to low-attention entities ($-$58\%; 95\% CI [$-$73\%,$-$38\%]).
Studying the heterogeneity of deplatforming, we also find that both permanent and temporary deplatforming significantly reduced online attention and that users banned for spreading misinformation seem to have their subsequent online attention reduced further than those banned for other reasons.

\xhdr{Implications}
Overall, we contribute to an emerging literature mapping the effectiveness of content moderation interventions and helping to guide platform governance practices (\eg, ~\cite{srinivasan_content_2019,jimenez-duran_economics_2023,clayton_real_2020,ecker_explicit_2010}).
Our results offer empirical evidence that sanctioning influencers significantly reduces online attention directed at them. Further, we find similar effects for both temporary and permanent deplatforming.
Therefore, platforms and other stakeholders should consider temporary bans to prevent harm caused by influencers' online presence, particularly those receiving widespread online attention.

\section{Background and Related Work}
\label{sec:rw}

\subsection{Background}

\xhdr{Reddit}
Reddit is a community-oriented social media platform centered around ``subreddits'' where users can contribute with posts and comments.
We use Reddit to obtain information about deplatforming events because the platform is commonly used to discuss internet-related events~\cite{leavitt2017upvote,roozenbeek2017read}.
Since Reddit users are mostly Western and English-speaking~\cite{roozenbeek2017read}, this choice biases our data to deplatforming events relevant to English-speaking countries.
Yet, it is worth noting that content moderation, in general, is biased toward Western and English-speaking countries~\cite{shahid2023decolonizing}.

\xhdr{Google Trends}
Google Trends is a freely accessible tool that allows anyone to measure the popularity of search queries on Google.com, the world’s largest search engine.
Queries can be specified as plain text or as identifiers from Google's knowledge graph~\cite{singhal2012introducing}, \eg, Alex Jones corresponds to the identifier /m/01\_6j\_.
The research community has extensively used Google Trends in diverse scenarios, from nowcasting economic indicators~\cite{choi2012predicting} to estimating the prevalence of diseases~\cite{nuti2014use}.
One issue with using Google Trends is that it yields rounded and normalized results, i.e., the output is quantized to integer precision and scaled such that the maximum value of every time series equals 100.
To address this issue, we leverage Google Trends Anchor Bank (G-TAB)~\cite{west2020calibration}, a method for calibrating Google Trends time series such that time series for an arbitrary number of Google queries can be expressed on a common scale with high resolution.

\xhdr{Wikipedia}
Wikipedia is the world's largest encyclopedia and one of the most visited sites on the Web.
People turn to Wikipedia for various information needs, from keeping up with current events to randomly surfing across the Web due to boredom~\cite{singer2017we}.
The Wikimedia Foundation makes aggregate statistics of Wikipedia page views publicly available, and the research community has largely used this data as a metric of public interest: from measuring public interest in biodiversity and conservation~\cite{mittermeier2021using} to forecasting election results~\cite{yasseri2016wikipedia}.
It is worth noting that viewing Google and Wikipedia as completely unrelated data sources is ill-informed. 
Previous research has shown that both platforms are interdependent, with the search engine heavily relying on Wikipedia to provide factual content~\cite{vincent2019measuring,mcmahon2017substantial} and the online encyclopedia receiving a substantial amount of its visitors from Google~\cite{piccardi2021value, mcmahon2017substantial}.

\subsection{Deplatforming and other content moderation interventions}

In 2015, Reddit, which previously had a \textit{laissez-faire} approach to content moderation, created an anti-harassment policy~\cite{redditah} that resulted in the banning of several subreddits, most prominently \texttt{r/CoonTown} and \texttt{r/FatPeopleHate}, subreddits known respectively for racist and fatphobic (and generally toxic) content~\cite{redditah2}.
In perhaps the first major study around deplatforming, Chandrasekharan et al. (2017) \cite{chandrasekharan2017you} analyzed the aftermath of such bans on Reddit, finding that they decreased subsequent activity and toxicity of members of the said communities.
Similarly, Jhaver et al. (2021)~\cite{jhaver2021evaluating} examined how supporters of three prominent influencers behaved following their Twitter ban. Their findings were consistent with the previous Reddit study.

While the previous papers indicate that deplatforming ``works'' within specific platforms, other research suggests that deplatforming may backfire or be ineffective.
The keyword here is \textit{resilience}: fringe communities, often associated with specific ideologies, play key roles in the lives of its members, and thus, in the aftermath of deplatforming, these members are persistent in regrouping and in finding alternative ways to congregate~\cite{vu2023no, johnson2019hidden}.
One mechanism by which deplatforming can ``backfire'' is in this regrouping process, \ie, online communities relocate to alternative, fringe platforms, where moderation standards are lower~\cite{horta2021platform}.
For example, members of \texttt{r/CoonTown} and \texttt{r/FatPeopleHate} migrated to alternative platforms, in particular, \texttt{voat.co}~\cite{newell2016user}, where users became more toxic~\cite{monti2023online}.
The reach of these new platforms is often reduced; there are fewer newcomers and fewer active users compared to mainstream platforms, but whether a smaller but more toxic community is better begs the question~\cite{horta2021platform, monti2023online}.
Besides Reddit, such ``backfire'' effects of deplatforming were also shown on Twitter, where studies matched banned Twitter users with accounts with the same handle on Gab.
These studies found that, after being banned on Twitter, users became more active and toxic in the alt-tech clone~\cite{ali2021understanding, mitts2021banned}.
Last, in many cases, deplatformed users migrate to less public-facing platforms, like Telegram, which makes tracking their activity harder. For example, Urman and Katz (2022) found explosive growth in the messaging platform that coincides with the mass bans of far-right actors on mainstream social media platforms~\cite{urman2022they}.

However, the migration of communities from mainstream to fringe platforms is not the only issue with deplatforming as a moderation strategy.
Russo et al. (2023) show that deplatforming creates dual-citizen, coactive users who participate in both mainstream and fringe platforms~\cite{russo2023spillover}. 
This finding is particularly relevant to mainstream platforms as it shows that antisocial behavior happening elsewhere can bleed back in, resonating with a body of work indicating the disproportional influence of fringe platforms in shaping our information ecosystem~\cite{zannettou2017web, zannettou2018origins}.
Further, recent work suggests that deplatforming an entire fringe platform may not be effective, given that the banning of Parler by Amazon AWS did not decrease Parler users' consumption of fringe platforms altogether~\cite{horta2023deplatforming,agarwal2022deplatforming}.
This heterogeneity in the effects of deplatforming also happens at the community level, with some types of communities (\eg, QAnon) being more resilient to deplatforming than others (\eg, \texttt{r/FatPeopleHate})~\cite{monti2023online}, and at the user-level, with some types of users (\eg, those that exhibit antisocial behavior) migrating to the alternative platform more often~\cite{russo2022understanding}.

Online platforms like Reddit, Twitter, and Facebook use a variety of content moderation strategies to enforce laws (\eg, copyright law) and their community guidelines (\eg, Reddit's anti-harassment policy~\cite{redditah}).
In that context, deplatforming is a very specific moderation intervention that should be considered among the plethora of other interventions carried out by online platforms.
For example, on Reddit, a prominent moderation intervention is the \textit{quarantining} of problematic communities.
When a community is quarantined, three things happen:
1) users must be logged in to Reddit to access it.
2) the community's homepage is prefaced with a warning, e.g., saying that the community ``Contains a high volume of information not supported by credible sources.''
3) the community stops appearing in Reddit search and homepage, meaning that users need to use its direct URL to access the community.
Previous work suggests that this intervention decreased activity in the targeted communities, particularly the influx of newcomers, yet there was no prolonged decrease in toxicity, racism, and misogyny within targeted communities~\cite{chandrasekharan_quarantined_2022,trujillo2022make}.

Despite the potential impact of platform-level moderation interventions like deplatforming and quarantining on our information ecosystem, most content moderation efforts focus on individual decisions regarding posts, comments, images, and videos regular users create. 
In that context, past research has analyzed the effect of content removals in various settings~\cite{srinivasan_content_2019,horta_ribeiro_automated_2023,jimenez-duran_economics_2023}, as well as the individual labeling of content that might be misleading and polarizing~\cite{bhuiyan_nudgecred_2021,clayton_real_2020,gao_label_2018,ling_learn_2023,zannettou_i_2021,porter_political_2022,mena_cleaning_2020,ecker_explicit_2010}.

\subsection{Relationship between prior and present work}

Our work differs from previous research in the nature and the scale of online traces analyzed.
Using Google Trends and Wikipedia Pageviews, publicly available signals not typically used in content moderation research, we study the effect of deplatforming in online traces capturing \emph{passive engagement} across over a hundred deplatforming events.
This allows us to assess the causal impact of deplatforming holistically but also to study the heterogeneity of the effect of the intervention (\ie, when does deplatforming work best?).
Also, we focus on the deplatforming of influencers, whereas much of previous research has studied the deplatforming of fringe communities~\cite{russo2022understanding,russo2023spillover,chandrasekharan2017you,trujillo_one_2023,trujillo2022make} or even of entire social media platforms~\cite{horta2023deplatforming}.

\section{Data Collection and Curation}

\begin{figure}
\centering
\includegraphics[scale=.9]{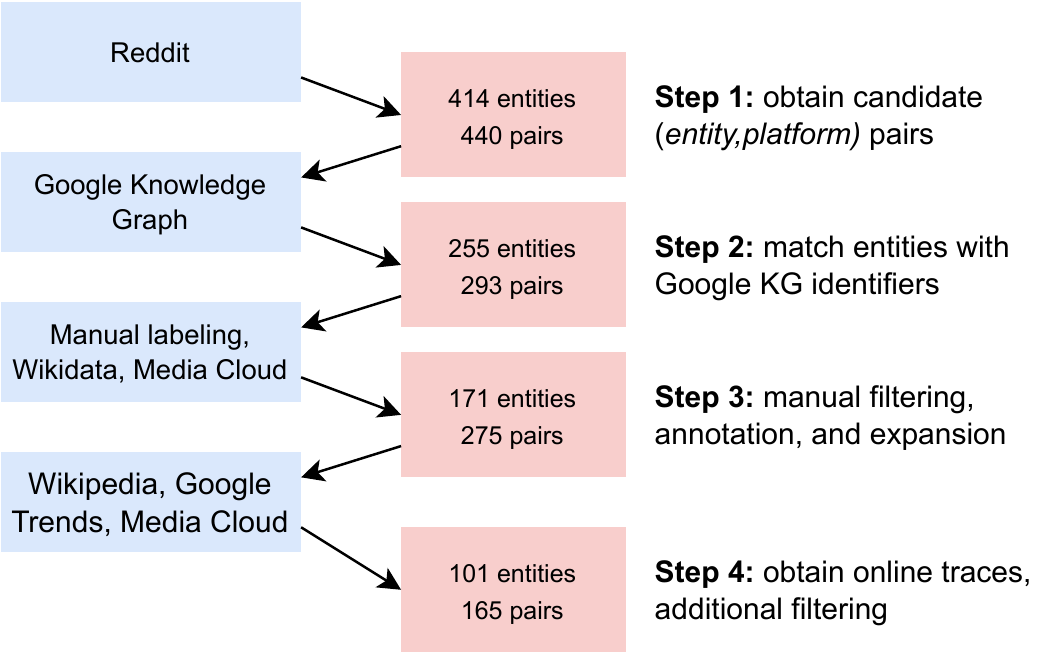}
\caption{
\textbf{Overview of our data collection and curation pipeline.}
Starting from Reddit, we obtain 440  $\langle$entity, platform$\rangle$ pairs, each corresponding to a deplatforming event (\emph{Step 1}).
Then, we link entities to Google Knowledge Graph identifiers, which are subsequently linked to GKG-ids (\emph{Step 2}) and manually filter, annotate, and expand the data (\emph{Step 3}).
Last, we obtain online traces corresponding to each entity from Wikipedia, Google Trends, and Media Cloud (a source of internet news), performing additional filtering to ascertain the quality of the online traces (\emph{Step 4}).
}
\label{fig:data_processing}
\end{figure}

We summarize our data collection and curation methodology in \Figref{fig:data_processing} and detail the steps below.

\subsubsection{Step 1: Obtaining candidate $\langle$entity, platform$\rangle$ pairs}
To obtain information about ``who was deplatformed where,'' we used a semi-automatic approach using Reddit data.
In Reddit, a large community-oriented social network, it is common for users to share news pieces and commentary on internet-related events, including when individuals, media outlets, or even entire online communities are banned from social media platforms.
Considering all Reddit posts until the end of August 2021, we extracted $\langle$entity, platform$\rangle$ pairs from the title of Reddit posts using a pattern-based bootstrapping methodology inspired by Quootstrap~\cite{pavllo2018quootstrap}.
Starting from the seed patterns ``$\langle$entity$\rangle$~(was | were | has been | $\varnothing$) banned from $\langle$platform$\rangle$,'' we iterated these two phases:
\begin{enumerate}
    \item \textbf{Pair extraction step:} Extract $\langle$entity, platform$\rangle$ pairs in the data that match  previously discovered patterns. 
    \item \textbf{Pattern extraction step:} Discover new patterns expressing the previously extracted $\langle$entity, platform$\rangle$ pairs.
\end{enumerate}
At the end of phase \#2, we manually filtered the set of extracted $\langle$entity, platform$\rangle$ pairs as well as the set of extracted patterns, removing incorrect pairs, as well as patterns that occur infrequently (as, in practice, we found that these patterns generalize poorly across iterations).%
\footnote{More precisely, we removed patterns that occurred fewer than 14 times. This threshold was determined by analyzing the distribution of pattern occurrences.}
Note that here, we considered as `incorrect' titles that did not indicate a deplatforming event, e.g., ``Activists push for Twitter to Ban Donald Trump.'' 
We repeated this iterative process twice, obtaining 440 $\langle$entity, platform$\rangle$ pairs involving 414 entities and 18 distinct platforms. The final list of patterns used is listed in \Tabref{tab:final_patterns} at the end of the paper.


\subsubsection{Step 2: Match entities with GKG-ids}

We then managed to link 255 of the extracted entities to Google Knowledge Graph identifiers
(henceforth referred to as GKG-ids; \eg, Alex Jones corresponds to the identifier {/m/01\_6j\_}%
\footnote{Note that GKG-ids are a superset of the identifiers of Freebase, a collaborative knowledge base acquired by Google in 2010. For the relationship between Freebase and Wikipedia, refer to Tanon et al.~\cite{tanon2016migration} }%
)
as well as Wikidata ids
(Wiki-ids; \eg, Alex Jones corresponds to the identifier {Q319121}). 
These identifiers helped us link entities in our dataset with their corresponding Google Trends and Wikipedia pageview data, respectively.

\subsubsection{Step 3: Manual filtering, annotation, and expansion} Next, we filtered, annotated, and expanded the data with the aid of Wikidata~\cite{vrandevcic2014wikidata}, Media Cloud~\cite{roberts2021media}, Wikipedia, and the search functionalities of Reddit and Google. This resulted in a dataset with 171 entities and 275 deplatforming events.

\begin{table}
\caption{Examples of sources obtained for $\langle$entity, platform$\rangle$ pairs.}
  \begin{minipage}{\textwidth}
  
\centering

\begin{tabular}{p{3cm}|p{3.25cm}|p{2.5cm}|p{3cm}}
\toprule
 \textbf{Date} & \textbf{Entity} &  \textbf{Platform} & \textbf{Source}\\ \midrule
2018-04-16 & Richard B. Spencer & Facebook & BBC\footnote{\tiny \url{https://www.bbc.com/news/technology-43784982}} \\
2021-06-04 & Yair Netanyahu & Twitter & Times of Israel\footnote{\tiny \url{https://www.timesofisrael.com/yair-netanyahu-temporarily-blocked-from-social-media-for-protest-call}}\\
2020-08-07 & Tommy Robinson & Instagram & Daily Caller\footnote{\tiny \url{http://dailycaller.com/2018/08/07/instagram-tommy-robinson-ban}} \\
\bottomrule
\end{tabular}
\end{minipage}
\label{tab:sources}
\end{table}

\xhdr{Ensuring the completeness of deplatforming events}
We search for additional deplatforming events for the entities already in the dataset using Media Cloud, an open-source collection of news on the Web~\cite{roberts2021media}.
For each entity, we conducted a search using all curated patterns obtained in Step 1.%
\footnote{For example, for Alex Jones, we used the query \textit{"bans Alex Jones" OR "suspends Alex Jones" OR "suspended Alex Jones" OR 
"banning Alex Jones" OR "Alex Jones locked out of" OR "Alex Jones blocked on" (...) has been banned from"}.}
With this procedure, we managed to retrieve 8036 stories about 124 of the entities.
For each entity, we manually inspected the news headlines (opening the URL), looking for different deplatforming events that were not covered in the original dataset.
We found nine additional deplatforming events corresponding to eight distinct entities. 
In many cases, those events were associated with entities that were banned from several distinct platforms, \eg, Alex Jones (banned from six platforms) and David Duke (banned from two platforms).
Overall, this step worked as a sanity check for the internal completeness of the data (\ie, for the entities considered, are all meaningful deplatforming instances represented?)\ but also allowed us to further refine our data through the addition of the nine deplatforming events we found.
We acknowledge that despite this rigorous effort, some salient deplatforming events might have escaped our data collection or might not have appeared in any news media stories at all. 
Therefore, we present our approach as our best effort at ensuring the completeness of prominent deplatforming events about relevant entities.

\xhdr{Ensuring the correctness of deplatforming events}
One author of this paper manually checked whether each $\langle$entity, platform$\rangle$ pair was correct (\ie, was the entity banned from the platform?).
They checked the links shared on Reddit for each pair to find a reliable source (\eg, mainstream news websites) confirming the deplatforming event.
They also searched for the entity on Google News and Wikipedia where necessary. 
Through this process, which took around 10 hours, they ensured that all deplatforming events were correct and attributed a deplatforming date to each event. 
We provide a handful of examples of sources and dates for different $\langle$entity, platform$\rangle$ pairs in \Tabref{tab:sources}, e.g., per BBC, Richard B. Spencer was banned from Facebook on the 16th of April, 2018.

\begin{table}
\centering
\caption{\textbf{Entity-level labels assigned to person entities along with examples.}Size here corresponds to the number of deplatforming events with each label  in the dataset obtained at the end of \textit{Step 3}.}
\small
\begin{tabular}{p{4cm}|p{9cm}}
\toprule
\textbf{Label} (size) &  \textbf{Examples}\\ \midrule
Politician ($n=31$) &  
David Duke,
Donald Trump,
Marjorie Taylor Greene,
Ron Paul
\\ \midrule
Internet personality   ($n=51$) & 
Blaire White
Carl Benjamin
Paul Joseph Watson
Stefan Molyneux \\ \midrule
Media personality   ($n=65$) &  Candace Owens,
Graham Linehan,
Katie Hopkins,
Tila Tequila  \\ \midrule
Fringe movement   ($n=83$) &  
James Allsup,
Gavin McInnes,
Nicholas J. Fuentes,
David Duke
\\
\bottomrule
\end{tabular}
\label{tab:entity-level}
\end{table}

\xhdr{Entity-level labels} 
We used Wikidata and retrieved whether each entity is an instance of either a ``Person'' (Human Q5; 146 entities in our data) or an ``Organization'' (Organization Q43229; 25 entities in our data).
We also considered organizations websites (Q35127; \eg, SciHub) and political movements (Q2738074; \eg The Boogaloo movement) aligned with Wikidata's definition~\cite{wikidatadef} of an {organization} as a {``social entity established to meet needs or pursue goals.''}
Then we annotated, for each \textit{person} in the dataset, whether they are 
1) a politician; 2) a media personality; 3) an Internet personality; or 4) associated with fringe movements and/or ideologies. 
Considering the first sentences in each person's article, these labels were assigned, which, according to Wikipedia's Style Manual~\cite{wikipediastylemanual}, should tell the nonspecialist reader what or who the subject is.
Note that labels are non-exclusive, e.g., Alex Jones has both the ``media personality'' and the ``fringe movement'' labels.
To validate our labels, two authors of this paper manually verified the four labels for 39 entities, reading their Wikipedia entries of each, reaching a consensus through discussion in cases of disagreement. They agreed on 96.7\% (151/156) of the labels assigned in this fashion, which we considered adequate for subsequent analyses (done with the labels assigned by a single coder). 
\Tabref{tab:entity-level} depicts the (non-exclusive) entity-level labels assigned to person entities.

\xhdr{Event-level labels} We developed a taxonomy for the reason an entity was suspended from the online platforms by examining Meta's community standards~\cite{facebookrules}, Twitter Rules~\cite{twitterrules}, and the headlines of all news sources obtained. Our final taxonomy, shown in \Tabref{tab:event-level} along with examples and relevant parts of the policy documents, consists of three categories, selected based on their prominence in platform policies and frequent occurrence in deplatforming-related news: 
A)~\emph{Hate, Harassment, Incitement to Violence}; 
B)~\emph{Misinformation, Platform Manipulation}; and 
C)~\emph{Other/Unknown}, a category encompassing less common reasons (\eg, lewd content, copyright infringement), as well as cases where the reason for the suspension was not clear. 
Further, we classified each deplatforming event as either ``permanent'' or ``temporary.'' We considered the bans \textit{permanent} when the social media pages for the entity in question were online at the time of annotation (October 2021) and temporary otherwise. Note this was before the wave of ``re-platforming'' events started in late 2022 on Twitter after Elon Musk's acquisition of the platform.
To validate our labels, another author manually verified the two labels for 30 entities, reaching 96.5\% agreement (56/58) with the original labels, which we considered adequate for subsequent analyses (again, with the labels from a single coder).

\begin{table}
\centering
\small
\caption{\textbf{Categorization of deplatforming reasons used in this paper.} We consider three categories (first column), pointing associated policies (second column), and example headlines (third column). Note that the associated policies can be found in Meta and Twitter respective websites~\cite{twitterrules, facebookrules} and that the headlines given as examples are from the sources extracted for each deplatforming instance. Size here corresponds to the number of deplatforming events with each label  in the dataset obtained at the end of \textit{Step 3}.}
\begin{tabular}{p{3.5cm}|p{3.75cm}|p{3.76cm}}
\toprule
\textbf{Reason} (size) &  \textbf{Associated Policies} & \textbf{Example Headlines} \\ \midrule
Hate, Harassment, Incitement~to Violence \newline ($n=166$) &
\textbf{Meta}: Hate Speech, Violence and Incitement, and Dangerous Individuals and Organizations Policies 
\newline  \newline 
\textbf{Twitter}: Abusive Behavior, Hateful Conduct and Violent Organizations Policies &
Twitter suspends Azealia Banks for transphobic tweets
\newline \newline  
YouTube removes 3 prominent white supremacist channels
\\ \midrule
Misinformation and Platform Manipulation \newline ($n=36$) &
\textbf{Meta}: False News, Manipulated Media and Inauthentic Behavior Policy.
\newline \newline
\textbf{Twitter}: Platform manipulation and spam, Civic integrity, Synthetic and manipulated media policy, COVID-19 misleading information policy
& 
Charlie Kirk: Trump supporter has Twitter account locked for spreading misinformation about mail-in voting
\newline \newline
Facebook suspends Cambridge Analytica (\dots) 
\newline \newline  
\\ \midrule
Other/Unknown \newline 
($n=73$) & 
This label was assigned to unclear cases or those that did not fit the above labels.
& 
Courtney Stodden, 17, Banned from Facebook for “Sexy” Shots
\newline \newline
Facebook blocks "Atheist Republic" on government directive
\\
\bottomrule
\end{tabular}
\label{tab:event-level}
\end{table}

\xhdr{A dataset of deplatforming events}
We obtain a carefully curated, comprehensive dataset of 275 deplatforming events between 2010 and August 2021. While we subsequently filter this data to answer the research questions at hand, e.g., keeping only entities for which we obtain meaningful online traces, we also make this intermediate data available. This data can be used by the research community to explore a wider variety of relevant questions about the dynamics of social media deplatforming, sanctioned influencers, and the linkages between social media and news media.

\subsubsection{Step 4: Obtain online traces, additional filtering}
Finally, we retrieved online attention data from Google Trends and Wikipedia and further filtered and processed the data to answer our specific research questions.

 \xhdr{Data crawling} 
 To obtain Wikipedia pageviews, we used a well-established API%
\footnote{Wikimedia's pageview API, \url{https://wikitech.wikimedia.org/wiki/Analytics/AQS/Pageviews}}
offered by Wikimedia.
To obtain search interest, we use Google Trends Anchor Bank (G-TAB), a method for calibrating Google Trends data that allows us to obtain queries on a universal scale and to minimize errors stemming from Google's rounding to the nearest integer~\cite{gtab}.
We considered only data between 1 April 2015 and 1 September 2022. As discussed in the following paragraph, we only consider bans after 2016 (the vast majority), and these dates correspond to the 
period from the first [last] ban minus [plus] 1 year. 
We search for entities using the GKG-ids (which are matched one-to-one to every Wikipedia page). 

\xhdr{Additional filtering} Starting from the 171 entities from \textit{Step 3}, 
we excluded 23 entities that we labeled as Organizations (as we focus on individuals) and 35 entities for which data was unavailable (or noisy%
\footnote{Even using G\hyp{}TAB, the signal provided by Google for entities with low attention remains unreliable, as Google adds random noise to the queries. This noise is not noteworthy for most queries, but for entities that receive low attention, it causes substantial fluctuations in value. In this context, we filtered entities with average attention smaller than 0.001 units.
}%
) for one of the two considered data sources. 
Further, to simplify our data analysis, we filtered eight entities that were banned before 2016 and 4 entities that were banned from platforms other than YouTube, Facebook, Instagram, or Twitter.

\begin{figure}
\centering
\begin{minipage}[t]{\textwidth}
\begin{minipage}[t]{0.49\textwidth}
\subcaption{Number of ban groups per merge threshold}
\label{fig:step4a}
\includegraphics[scale=0.5]{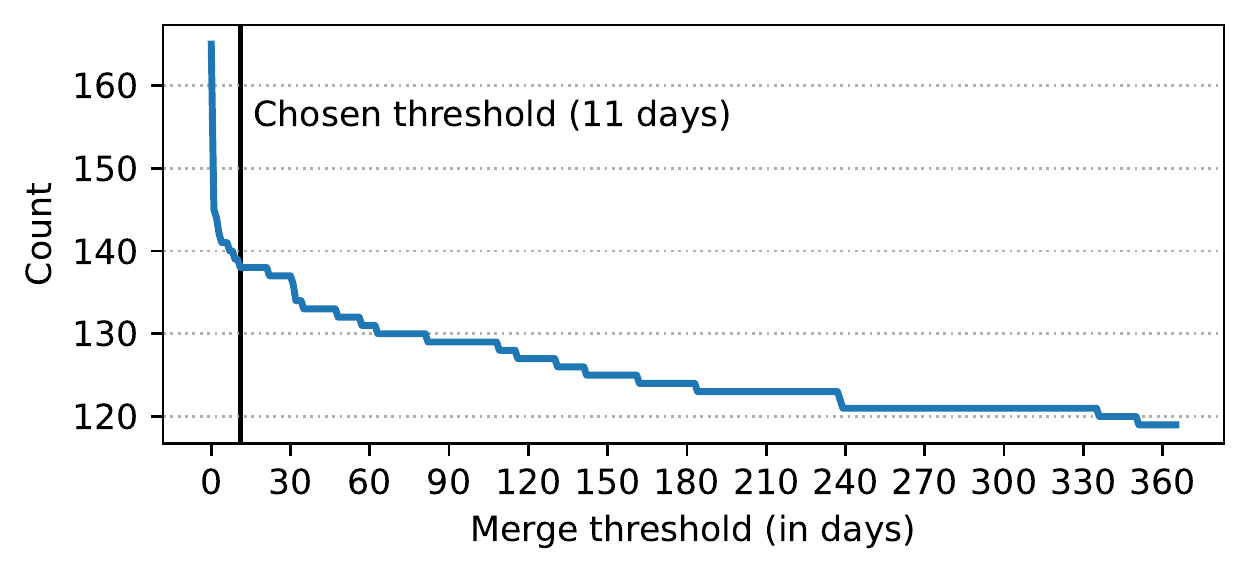}
\end{minipage}\hfill%
\begin{minipage}[t]{0.49\textwidth}
\subcaption{Number of bans per month}
\label{fig:step4c}
\includegraphics[scale=0.5]{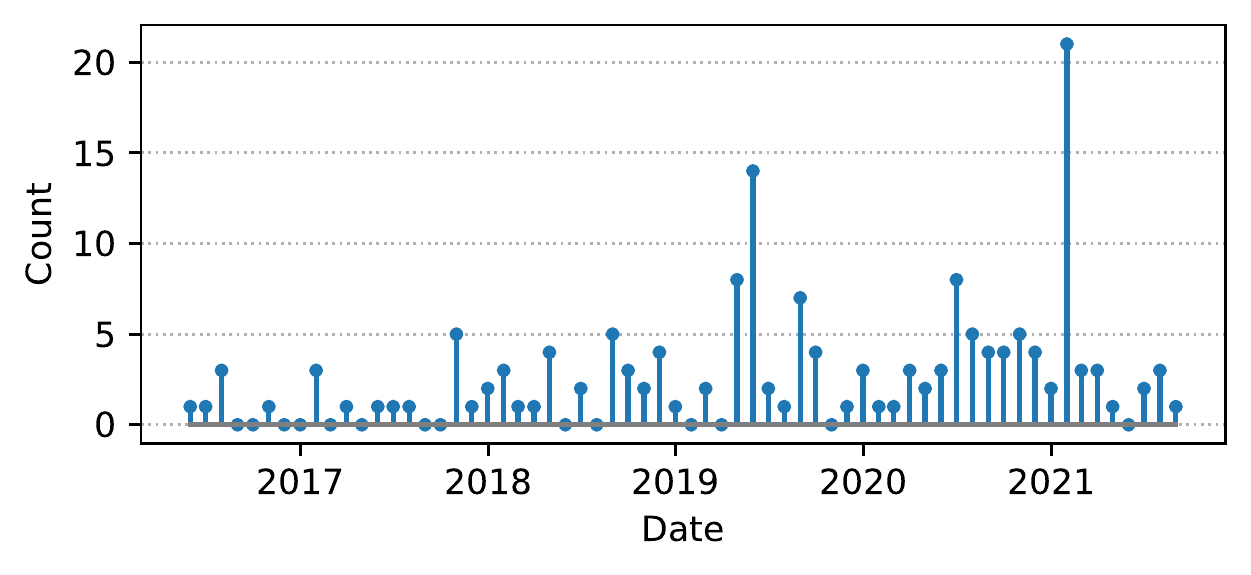}
\end{minipage}
\end{minipage}

\begin{minipage}[t]{\textwidth}
\begin{minipage}[t]{0.49\textwidth}
\centering
\footnotesize
\subcaption{Event-level labels}
\label{fig:step4b}
\vspace{0.875mm}
\begin{tabular}{p{1.25cm}|p{3.25cm}|p{0.4cm}}
\toprule
\textbf{Label} & \textbf{Category} & $n$ \\ \midrule
Reason & Hate/Harassment/Incitement & 116 \\
& Manipulation/Misinformation & 23 \\
& Other/Unknown & 26 \\ \midrule
Temporary & True & 55\\
& False & 110 \\  \bottomrule
\end{tabular}
\end{minipage}\hfill%
\begin{minipage}[t]{0.49\textwidth}
\centering
\footnotesize
\subcaption{Entity-level labels}
\label{fig:step4d}
\vspace{0.875mm}
\begin{tabular}{p{1.25cm}|p{3.25cm}|p{0.4cm}}
\toprule
\textbf{Label} & \textbf{Category} & $n$ \\ \midrule
Type & Politician & 28 \\
 & Media personality & 33 \\
 & Internet personality & 40 \\
 & Fringe movements & 68 \\ \bottomrule
\end{tabular}
\end{minipage}
\end{minipage}

\caption{
\textbf{Dataset details.}
\textit{(a)} We show how the number of ban groups varies when changing the \textit{``merge threshold,''} \ie, the maximum number of days between two bans in the same ban group. We also show the chosen threshold. 
\textit{(b)} We depict the number of bans per month in the final dataset. 
\textit{(c)/(d)} Number of deplatforming events associated with each entity\hyp{} and event-level labels in the final dataset.}
\label{fig:step4}
\end{figure}

\xhdr{Ban groups} 
Several platforms often ban individuals within a couple of days for the same reason, \eg,  on 6 August, 2018, YouTube, Facebook, and Instagram banned Alex Jones simultaneously for his promotion of conspiracy theories~\cite{ajban}. 
Then, exactly one month later, on 6 September, Twitter also banned Jones due to his behavior on the platform~\cite{ajban2}. 
To simplify our analyses, we introduce the concept of a ``ban group,'' a series of highly connected bans that are close in time. For example, we determined that Jones's first three bans (YouTube, Facebook, and Instagram) should be in the same ban group, while the Twitter ban should not.
Using the elbow method (\cf\ \Figref{fig:step4a}), we determined that all bans imposed within 11 days are in the same ``ban group.'' Manually inspecting the ban groups, we find that all bans grouped this way are highly related, \ie, relating to the same incident.

\xhdr{A dataset of deplatforming events and online attention data} In sum, at the end of this step, we obtain a dataset containing \numDeplatformedinfluencers person entities involved in \numDeplatformedEvents deplatforming events between January 2016 and September 2021, each linked to online attention traces from two different sources. Online traces are almost complete, with less than 5\% of data points missing around the 12 months of each deplatforming event considered (due to them being unavailable on either Google Trends or Wikipedia).
We plot the distribution of bans over the considered period in \Figref{fig:step4a} and provide the number of entity\hyp{} and event\hyp{}level labels in \Tabref{fig:step4b} and \Tabref{fig:step4d}, respectively.

\begin{figure}
\begin{minipage}{0.49\textwidth}
\centering
\includegraphics[scale=0.5]{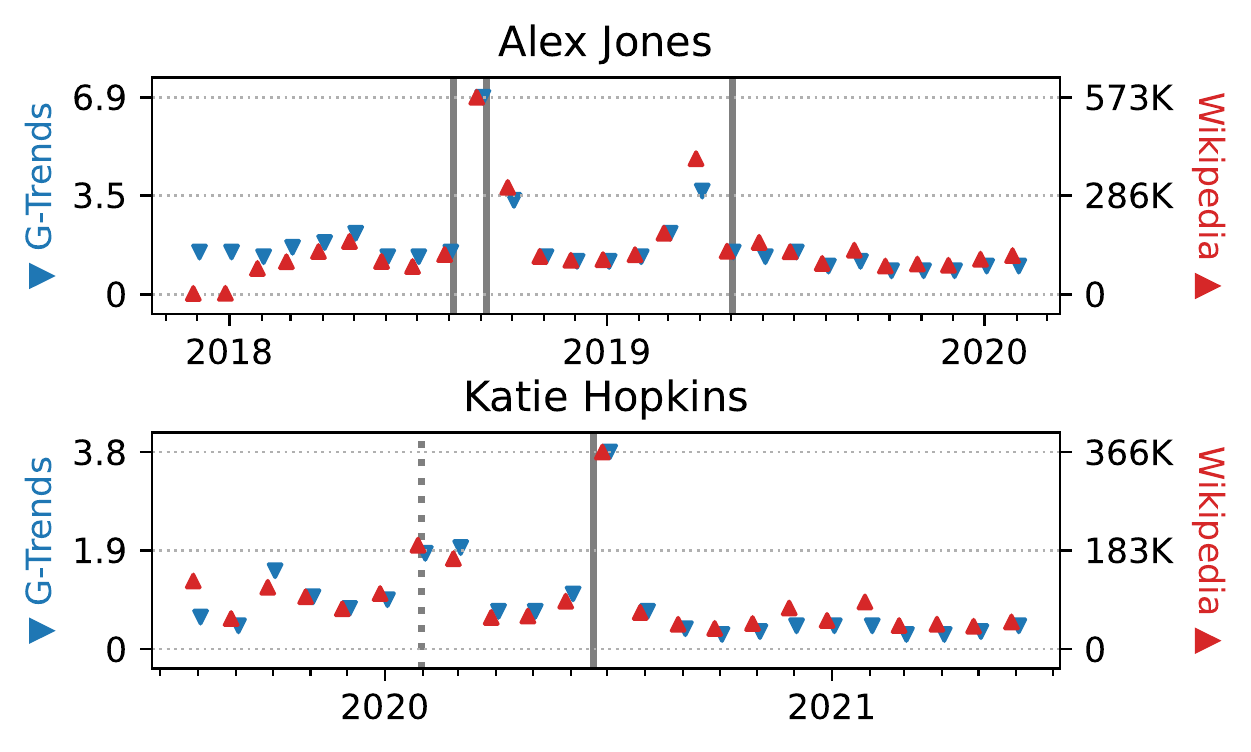}
\subcaption{}
\label{fig:ex_data}

\end{minipage}
\begin{minipage}{0.49\textwidth}
\centering
\includegraphics[scale=0.5]{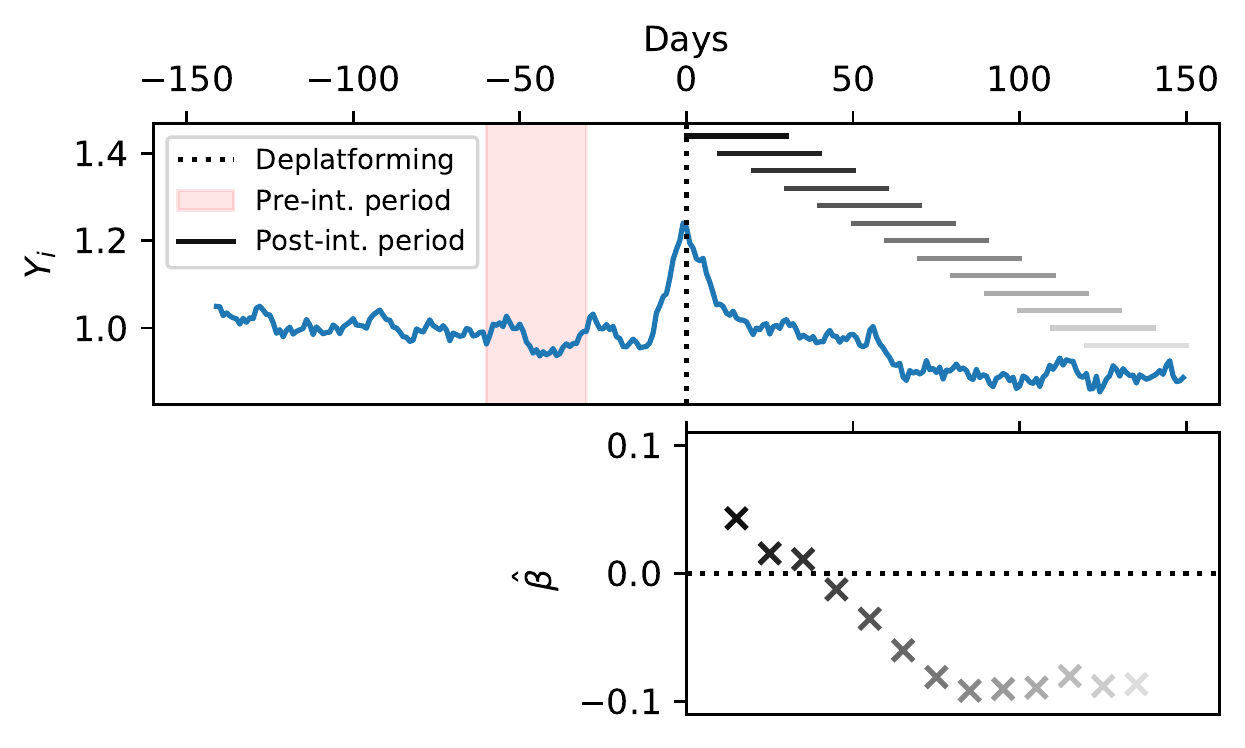}
\subcaption{}
\label{fig:ex_methods}

\end{minipage}
\begin{minipage}{\textwidth}
\caption{
\textbf{Illustration of our descriptive approach.}
(a) Monthly online attention traces for Alex Jones and Katie Hopkins (Google Trends in blue, left $y$-axis; Wikipedia pageviews in red, right $y$-axis). Vertical lines correspond to ``ban groups,''
\eg, in August 2018, several platforms banned Jones,%
and all those events are merged into the same ban group.
Dashed lines indicate that the suspension was temporary, \eg, Katie Hopkins was suspended temporarily from Twitter in late January 2020.
(b) Synthetic data depicting our fixed-effects approach, considering the online attention received by an entity $i$ ($Y_i$, $y$-axis top plot) deplatformed on day 0 ($x$-axis), we estimate the change in online attention post-deplatforming ($\beta$) as the average difference in attention ($Y_i$) between the post\hyp{} and pre\hyp{}intervention periods. When estimating Eq.~\eqref{eq:1}, we vary the post\hyp{}intervention period considered, as indicated by horizontal lines, to obtain time-varying estimates (bottom).
}
\end{minipage}
\end{figure}

\section{Changes in online attention following deplatforming}
\label{sec:pre-post}

We show the data collected for two entities, Alex Jones and Katie Hopkins, in \Figref{fig:ex_data}, illustrating the challenges associated with estimating the causal effect of deplatforming on online attention. 
Deplatforming is often preceded by a controversial event that boosts online attention toward that entity, and the intervention itself triggers subsequent attention toward the entity. In other words, the controversial event may explain an increase in online attention without a causal effect of deplatforming on online attention. This can be seen in the figure showing the sharp increases in online attention right before and right after the deplatforming event occurred.

Thus, before dwelling upon the causal question, we first focus on describing the changes using extensions of this simple fixed-effects model:
\begin{equation}\label{eq:1}
    Y_{i,t} = \beta \cdot 1\{D_{i,t} = 1\}  + \alpha_i + u_{i,t},
\end{equation}
where $Y_{i,t}$ is the natural logarithm of the attention toward entity $i$ at time $t$, $1\{D_{i,t} = 1\}$ is an indicator that equals 1 if $i$ was already deplatformed at time $t$ (and 0 otherwise), $\alpha_i$ is an entity-level fixed effect, and $u_{i,j}$ is the error term. 
Note that when estimating the model with least squares, $\beta$ is the average entity-level change in attention pre- \vs\ post-deplatforming.

Importantly, we do not fit this model using all of our data, but instead, consider a fixed pre\hyp{}intervention period and fit the regression model repeatedly considering different post\hyp{}intervention periods (Gelman's ``secret weapon''~\cite{gelman}). We illustrate this approach with synthetic data in \Figref{fig:ex_methods}. 
Given a deplatforming event that takes place on day 0, we consider the attention the entity received between days $-60$ and $-30$ as the pre-intervention period%
\footnote{We do not use the period right before the deplatforming event as a pre-intervention to prevent considering attention traces contaminated by a controversial event that has led to the entity's deplatforming.}
and estimate the model repeatedly for different post-intervention periods, indicated by horizontal lines in grayscale.
\Figref{fig:ex_methods} (bottom) shows the estimated coefficients $\hat{\beta}$ using different post-intervention periods.
This approach helps us estimate the effects of deplatforming at varying temporal distances.

\begin{figure}
\includegraphics[scale=0.5]{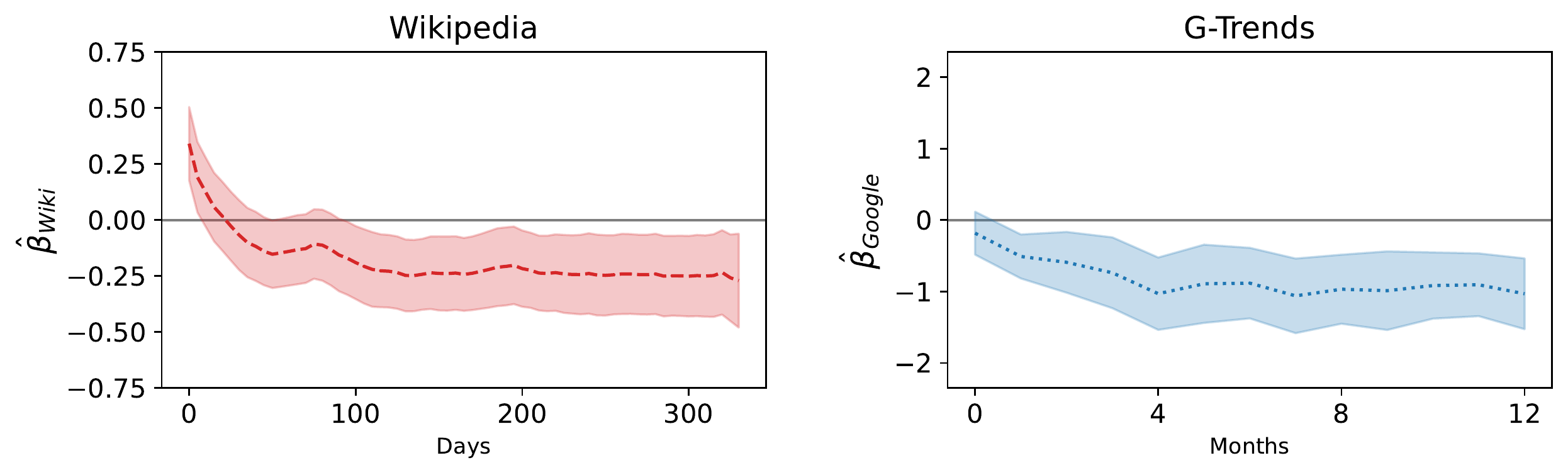}

\caption{
\textbf{Results from fixed-effects model.}
We show the estimated parameters of our fixed-effects models varying the start date of the 30-day post-intervention period considered for Wikipedia pageviews (left) and Google Trends (right). 
Specifically, we show the estimated $\beta$ for the model depicted in Eq.\eqref{eq:1}, which captures the average difference in attention pre\hyp{} \vs post\hyp{}deplatforming for the last deplatforming event of each entity in our datasets.}
\label{fig:me1}
\end{figure}

One practical challenge with naively using the model in Eq. ~\eqref{eq:1} to estimate changes in online attention post-deplatforming is that, as shown in \Figref{fig:ex_data}, entities can be deplatformed multiple times. 
Thus, the post-deplatforming periods used in estimating the model can coincide with subsequent deplatforming events. 
To prevent this issue, we estimate the model using the last deplatforming event and discarding entities who happened to have another deplatforming event in the pre-intervention period.%
\footnote{We also estimated the model using only the \textit{first} deplatforming event. This analysis yields similar results but is much noisier, as the number of units used to estimate the model varies with the period considered.}
Further, note that observations within each entity are likely to be serially correlated, which makes the estimates of $\text{Var}(\hat{\beta})$ inconsistent under the standard OLS estimator. We hence estimate standard errors using robust standard errors clustered at the entity level~\cite{cameron2015practitioner} to prevent this, obtaining consistent, albeit conservative, estimates of $\text{Var}(\hat{\beta})$.

We depict the estimates of $\beta$ in \Figref{fig:me1}, considering the logarithmic number of Wikipedia pageviews (left) and the logarithmic G-TAB estimate (right) as the outcomes~$Y$. We use the setup illustrated in \Figref{fig:ex_methods}, considering days $-60$ to $-30$ (relative to the last deplatforming event of each entity, \ie, day 0, when deplatforming happened) as the pre-intervention period, and varying the post-intervention period in the 360 days (or 12 months) that followed the last deplatforming event, always considering a 30-day window.
For instance, when considering the estimate corresponding to day 100 in the plot, we are thus considering the model fit with attention data from dates $t\in [-60,-30) \cup [100, 130)$.%
\footnote{For Google Trends, data is aggregated in the monthly granularity, and thus we use month -1 as the pre-deplatforming period and estimate the model with data for each month after the moderation decision took place.}
Note that taking logarithms of the outcome prior to estimating the model in Eq.~\eqref{eq:1} diminishes the influence of entities that receive a lot of attention and allows us to interpret the estimated coefficients $\hat{\beta}$ as multiplicative; \eg, if $\hat{\beta} = 0.1$, this means that deplatforming led to roughly a 10\% increase in attention received ($e^{0.1} - 1 \simeq 10.5\%$). 

Analyzing \Figref{fig:me1}, we can conclude that deplatforming was correlated with significant decreases in attention considering both Google Trends and Wikipedia page views. For Wikipedia, in particular,  the attention received by deplatformed entities is initially positive compared with the pre-intervention period, becoming negative as time passes and remaining negative long after. 

\begin{figure}[t]
\centering
\begin{subfigure}[b]{0.49\textwidth}
\includegraphics[width=\linewidth]{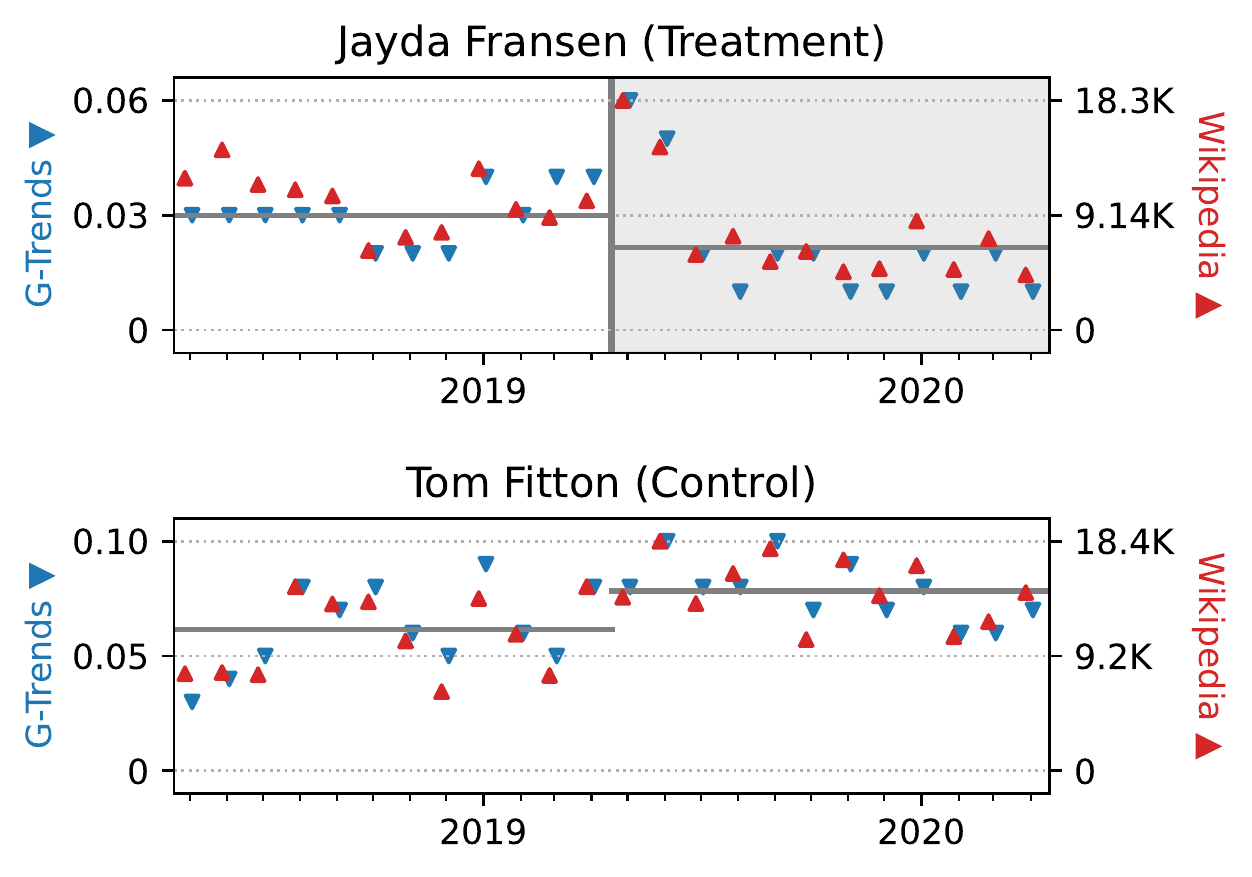}
\subcaption{}
\label{fig:didex1}
\end{subfigure}~
\begin{subfigure}[b]{0.49\textwidth}
\includegraphics[width=\linewidth]{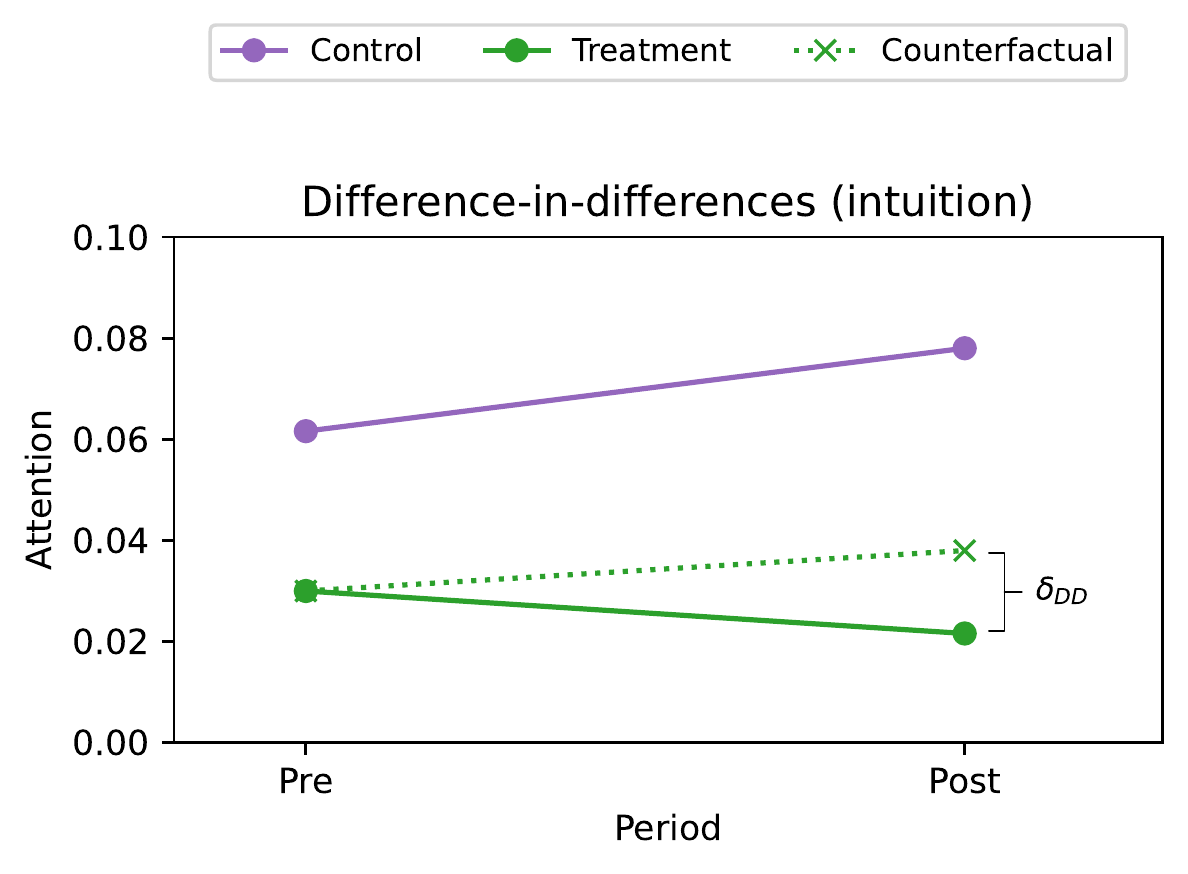}
\subcaption{}
\label{fig:didex2}
\end{subfigure}~

\caption{
\textbf{Illustration of our difference-in-differences (DiD) approach. }
We compare deplatformed and yet-to-be-deplatformed units, \eg, Tom Fitton \vs Jayda Fransen in (a), and estimate the effect of deplatforming with the difference-in-differences estimator (b). 
Note that in (a), horizontal lines represent the average attention received by the entities pre\hyp{} and post\hyp{}deplatforming. 
These values are also represented in (b) with circles.
The DiD estimator $\delta_{DD}$ constructs a counterfactual estimate of how the outcome of the treatment group would have progressed had the group not received the treatment ($\times$ in the plot), contrasting it with how the outcome actually changed between the pre and post periods.
}

\label{fig:examples_causal}
\end{figure}

\section{The causal effect of deplatforming on online attention}
\label{sec:did}
In this section, we estimate a lower bound of the causal effect of deplatforming on online attention.
One notable shortcoming of the ``pre \vs\ post'' analyses performed in \Secref{sec:pre-post} is that they do not account for potential trends in outcomes.
If online attention toward all entities decreases, we could find a negative decrease in post-deplatforming attention even without an effect.
We address this issue with a difference-in-differences (DiD) approach (further explained below).
The other substantial shortcoming is that, as mentioned in \Secref{sec:pre-post}, the event that caused deplatforming may also impact online attention, acting as a confounder. While the DiD methodology used here does not, by itself, solve this second issue, we argue that, as this confounding event is only likely to \textit{increase} the subsequent online attention, this makes the estimate obtained by the DiD estimator a lower bound (see  \Secref{sec:discussion} for details).

Our difference-in-differences approach compares deplatformed units and yet-to-be-deplatformed units,%
\footnote{We do not consider ``never-deplatformed'' control influencers as it is unclear whether they would be a meaningful control group or even how we obtain these influencers to begin with.}
e.g., in \Figref{fig:didex1}, we show the online attention received by Jayda Fransen and Tom Fitton; given that Tom was deplatformed many months after Jayda, he will be used as a ``control'' when considering Jayda's deplatforming. 
The gist of the identification strategy is the parallel-trends assumption: if, in the absence of treatment, the difference between the ``treatment'' and ``control'' groups is constant over time, then we can estimate the causal impact of deplatforming with the difference-in-differences estimator. We illustrate this in \Figref{fig:didex2}, considering Jayda and Tom's time series. We calculate the pre \vs post difference in online attention for the control group and assume that the treatment group would behave similarly, obtaining a counterfactual estimate for the treatment group, indicated by an orange cross and a dashed orange line. Then, we estimate the causal effect as the difference between how much the treatment group actually changed and the counterfactual estimate ($\hat{\delta}_{DD}$ in \Figref{fig:didex2}).

While the intuition behind parallel trends is the clearest in the simple ``2 by 2'' difference-in-difference illustrated in \Figref{fig:didex2}, observational studies often opt for a ``leads-and-lags'' difference-in-differences specification~\cite{cengiz2019effect}:
\begin{equation}
\label{eq:dd2}
    Y_{i,t} = \alpha_i + \lambda_t + 
    \sum_{\ell < -3} \delta_{\ell}  D^\ell_{i,t} + 
    \sum_{\ell \geq -2} \delta_{\ell}  D^\ell_{i,t} + 
    \epsilon_{i,t},
\end{equation}
where $Y_{i,t}$ is the outcome associated with unit $i$ at time $t$, $\alpha_i$ and $\lambda_t$ are unit and time fixed effects,  and $D^\ell_{i,t}$ is an indicator variable for unit $i$ being $\ell$ periods away from when it received the treatment.
The advantage here is that the $\delta_{\ell}$ coefficients obtained by estimating this model capture the causal effect $\ell$ months after the intervention, enabling the analysis of how the effect progresses with time.
Moreover, we also obtain coefficients associated with periods \textit{prior} to the intervention, \eg, $\delta_{-2}$. These coefficients allow for ``pre-testing,'' i.e., checking if the parallel trends assumption holds for periods before the intervention, like $\ell=-2$. This sanity check increases the credibility of the parallel trends assumption for the periods when the intervention actually happened (which is impossible to test).

However, this estimator is biased when we have multiple treatment periods because, under the hood, the fixed effect estimator erroneously compares newly treated units relative to already treated units~\cite{goodman2021difference}.
To address that, we use the ``stacked'' difference-in-differences approach proposed by Cengiz et al.~\cite{cengiz2019effect}. 
Their approach consists of considering an event window (here, we use 12 months before and after the ban) and then creating ``sub-datasets,'' each containing a treated unit and control units that have not yet been treated during the event window. 
Using the same ``leads-and-lags'' estimator in this stacked dataset yields an unbiased estimate, as the ``forbidden comparisons'' are pruned from the data.

\begin{figure}
    \centering
    \includegraphics[scale=0.5]{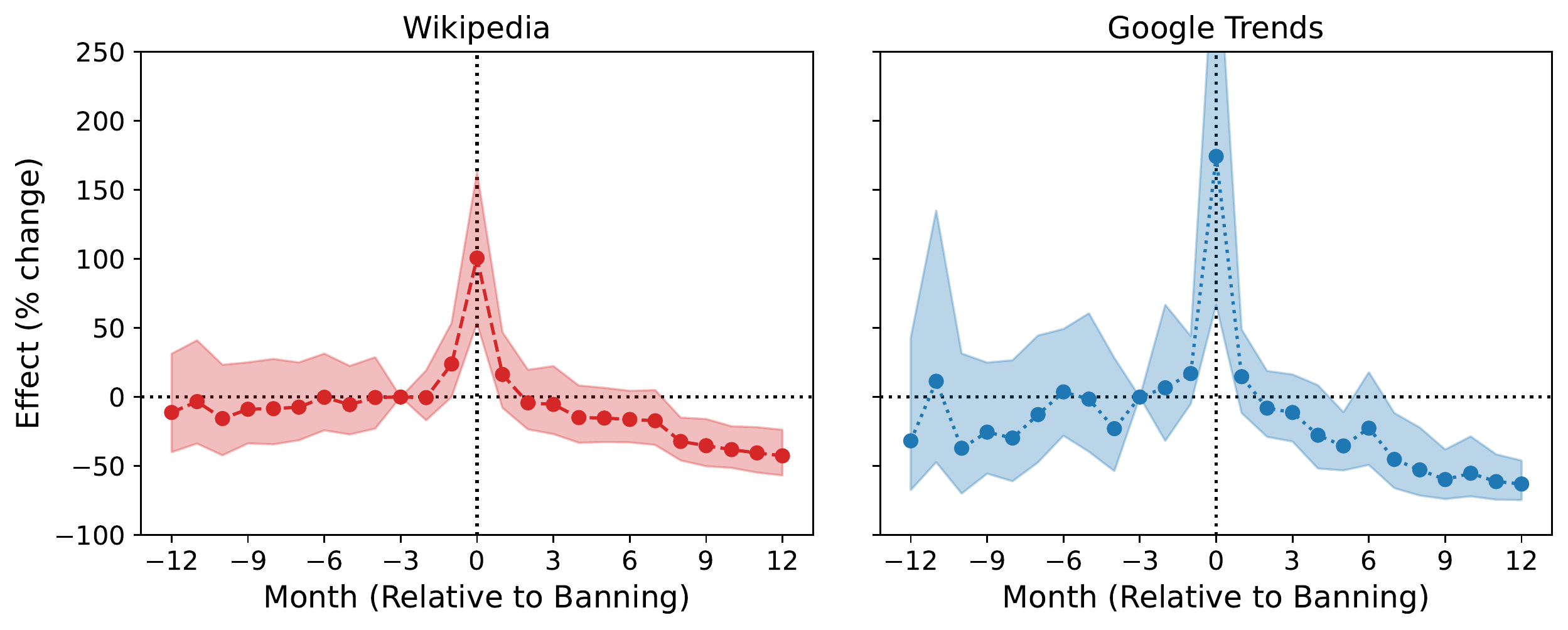}
    \caption{\textbf{Results from our difference-in-differences (DiD) approach.} We show the estimated reduction in online attention received by influencers due to deplatforming on Wikipedia (left) and Google Trends (right). We show the effect estimated across different months. Overall, these results indicate that deplatforming reduces online attention toward influencers.}
    \label{fig:did-res}
\end{figure}

\xhdr{Effect of deplatforming}
We show the estimated effects of deplatforming in \Figref{fig:did-res} for both Wikipedia pageviews (left) and Google Trends (right). Again, we consider logarithmic outcomes in the difference-in-differences model to interpret the effects multiplicatively. However, for simplicity's sake, here we show the percentage changes on the $y$-axis (\ie, $e^{\hat \beta}-1$ for a logarithmic effect $\hat \beta$) rather than the logarithmic effects.

Results with this quasi-experimental methodology are very similar to those obtained with the simple fixed-effects model. 
There is a sudden increase in online attention in the month when entities are deplatformed (Month = 0 in \Figref{fig:did-res}), followed by a continuous decrease in the attention that treated entities receive relative to control units.
After 12 months, we estimate that online attention towards deplatformed influencers is reduced by
$-$63\% (95\% CI [$-$75\%,$-$46\%]) on Google and by 
$-$43\% (95\% CI [$-$57\%,$-$24\%]) on Wikipedia.

We also show the estimated effect in the 12 months \textit{before} deplatforming, which allows us to assess the plausibility of the parallel trends hypothesis.
We find that trends remain parallel until right before the deplatforming event (Month $=-1$), when online attention spikes, suggesting that the event or events that triggered deplatforming also boosted online attention (see Section~\ref{sec:discussion} for further discussion).
\begin{table}[t]
    \centering
    \small
    \caption{\textbf{Heterogeneity of the effect.} We study how the effect of deplatforming varies depending on the characteristics of the deplatforming event (e.g., whether it is temporary) and on the characteristics of the influencer being deplatformed (e.g., whether it is a politician).
    Note that this is not the \textit{overall effect} for deplatforming events with these characteristics, but rather, the extent to which the effect of deplatforming events associated with these characteristics differs from the effect of deplatforming events that are not.
    }
    \input{tableeff}
    \label{tab:ff}
\end{table}

\xhdr{Heterogeneity of the effect} 
We adapt our difference-in-differences methodology to estimate whether the effect is heterogeneous across various dimensions. 
In particular, we decompose the effect by adding interactions with other variables and then analyze the coefficients associated with these interactions. 
For example, to measure whether permanent bans decrease subsequent online attention more than temporary bans, we consider the modified leads-and-lags specification:
\begin{equation}
\label{eq:dd3}
    Y_{i,t} = \alpha_i + \lambda_t + 
    \sum_{\ell < -2} \delta_{\ell}  D^\ell_{i,t} + 
    \sum_{\ell < -2} \gamma_{\ell}  D^\ell_{i,t} P_{i} + 
    \sum_{\ell \geq 0} \delta_{\ell}  D^\ell_{i} + 
    \sum_{\ell \geq 0} \gamma_{\ell}  D^\ell_{i,t} P_{i} + 
    \epsilon_{i,t},
\end{equation}
where $P_i$ is an indicator variable that equals $1$ only when the ban is permanent. 
This allows us to measure if the effect is significantly smaller or bigger in this subset of deplatforming events by analyzing the coefficients $\gamma$. (Note that the effect on influencers banned temporarily is captured by $\delta$, whereas the effect on the permanently banned is captured by $\delta + \gamma$.)

In this fashion, we consider four different models, each including different interactions. These models allow us to isolate whether platforming is more or less effective when considering
1) whether bans are temporary;
2) the reason for banning;
3) entity-level labels describing the influencers; and
4) the attention that the entities received in the pre-deplatforming period. In the latter case, we sort entities into two groups, according to the attention they received in the 12th month before deplatforming: high attention (top 1/3rd of influencers ranked per attention) \vs low attention (bottom 2/3rds).%
\footnote{High-attention influencers received, on average, 12${,}$977 pageviews per day in the 12th month before deplatforming, whereas low-attention influencers received 100.}

We show the coefficients associated with the interactions in \Tabref{tab:ff}.
Our key findings are threefold.
First, we find that temporary bans do not significantly differ from permanent ones at a 0.05 significance level (Model 1). This is true both on Google Trends ($p=0.739$) and Wikipedia ($p=0.791$).
Second, we find that the effects of deplatforming are significantly stronger for entities associated with higher attention in the pre-deplatforming period, both on Google Trends ($-$58.8\% 95\% CI [$-$72.6\%, $-$37.9\%]) and Wikipedia ($-$33.7\% 95\% CI [$-$48.2\%, $-$15.1\%]) (Model 2).
Third, we find that entities deplatformed because they spread misinformation suffer significantly larger decreases in online attention, again both on Google Trends ($-$59.8\% 95\% CI [$-$82.4\%, $-$7.7\%]) and Wikipedia ($-$48.7\% 95\% CI [$-$67.5\%, $-$19.2\%]) (Model 2).
We do not find systematic differences associated with the entity labels (Model 4).

A noteworthy methodological concern related to Model 2, where we study differences in the effect for high \vs low-attention influencers, is that we might observe a regression to the mean. In other words, we may find the decrease particularly salient for high-attention influencers because we filtered them to be high-attention before the intervention, selecting influencers with an unusually high-attention month.
However, we circumvent this issue by selecting high-attention users in a month far removed from the intervention (12 months before). Implicit in our DiD specification in Eq.~\eqref{eq:dd2} is that the reference month is set to $-3$, \ie, we compare the effect relative to three months before the intervention.%
\footnote{Other reference months, \eg, $-2$, yield similar results.}
This is long after the month used for selecting high-attention influencers, and, therefore, any ``regression to the mean'' would likely have already occurred (and not influence the post-intervention results, as we compare the online attention after the intervention with the online attention three months before the intervention.

\section{Discussion}
\label{sec:discussion}

In this section, we discuss the implications of our findings, methodological caveats, and promising venues for future work.

\xhdr{Deplatforming decreases online attention}
Our quasi-experimental analysis of \numDeplatformedEvents deplatforming events associated with \numDeplatformedinfluencers influencers indicates that deplatforming decreases online attention toward influencers. 
This key finding largely aligns with previous work on the deplatforming of influencers on Twitter~\cite{jhaver2021evaluating}. Yet, we argue that evidence from this previous study is, by itself, inconclusive due to the scope of the study (few events, one platform, active engagement only). 
In that context, we expect that our analysis gives stakeholders (including platforms and legislators) a more comprehensive and nuanced understanding of deplatforming.

\xhdrNoPeriod{How and when to deplatform?}
Deplatforming can be enacted in different ways (\eg, temporary or permanent); upon different kinds of influencers (\eg, politicians or media personalities); and for different reasons (\eg, hate speech or misinformation). Unlike previous work, \eg, \cite{jhaver2021evaluating,horta2023deplatforming,trujillo2022make}, we consider many deplatforming events, which allow us to capture variations in the effectiveness of the intervention.
We find that deplatforming is more effective when targeting popular influencers disseminating misinformation; yet, we argue that the key policy guidance our paper can provide comes from a null result: that temporary deplatforming was similar to permanent deplatforming in how it reduced online attention toward influencers.
We speculate that this may be partially due to deterrence, \ie, temporarily banned influencers may avoid rule-breaking to prevent harsher sanctions. Here we did not discriminate between short (\eg, one week) \vs long (\eg, one month) temporary bans, which could be an interesting venue for future work.
Broadly, permanent bans of prominent influencers have caused controversy as: 
the rationale behind the bans is often muddy~\cite{coaston_youtube_2018}; there is no clear ``reinstatement'' procedure for banned accounts~\cite{facebook_oversight_board_oversight_2021}; it is unclear if social media companies should be the ones to regulate speech~\cite{nunziato2022protecting}.
Having bans of limited time is aligned with the broader call for a more transparent~\cite{suzor2019we}, consistent~\cite{langvardt2017regulating}, and accountable~\cite{gorwa2020algorithmic} approach to social media moderation.

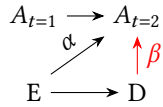
\begin{figure}[t]
    \centering
\begin{tikzpicture}
\node (1) {$A_{t=1}$};
\node [right = of 1] (2) {$A_{t=2}$};
\node [below = of 2] (3) {D};
\node [below = of 1] (4) {E};

\draw[-{Stealth}] (1.east) -- (2.west) ;
\draw[-{Stealth}] (4.north east) --  node[midway, above, sloped]  {$\alpha$}   (2.south west) ;
\draw[-{Stealth}] (4.east) -- (3.west) ;
\draw[red,-{Stealth}] (3.north) -- node[right] {$\beta$}  (2.south) ;
\end{tikzpicture} 
\caption{Directed Acyclic Graph depicting online attention before ($A_{t=1}$) and after ($A_{t=2}$) deplatforming ($D$). External events ($E$) may cause both the deplatforming of an entity and increase their online attention.
}
\label{fig:dag_disc}
\end{figure}

\xhdr{External events causing changes to attention and the deplatforming}
In a prominent deplatforming event, Twitter, Facebook, YouTube, and Instagram banned former US President Donald Trump's account shortly after the 6 January 2021 U.S.\ Capitol Attack, alleging that he was inciting violence. 
Here, it is likely that the online attention received by Trump after his deplatforming would be influenced by both his suspension from major platforms \textit{and} by the Capitol attack and his reaction to it.
This illustrates a problem faced by our analysis, as we want to estimate the effect of deplatforming, not of the event that triggered the deplatforming.
We argue this issue does not threaten the validity of our findings, given that we estimate that deplatforming harmed, rather than boosted, subsequent online attention.
This is because a potential confounding event is only likely to \textit{increase}  online attention post-deplatforming; and since the effect that we estimated with our model is negative, this confounder can only weaken the effect observed. 
Therefore, the results obtained here can be interpreted as a lower bound of the true effect of deplatforming.
In other words, if we could account for confounding due to an external event, the estimated effect of deplatforming would be even more strongly negative than the negative effects estimated by our model.

We more sharply illustrate this scenario with a directed acyclic graph as depicted in \Figref{fig:dag_disc}.
This simple model considers the online attention $A$ that an entity receives in two time periods $t\in\{1,2\}$, and the attention in $t=2$ is caused by 
1) how much attention the entity received in $t=1$,
2) the events $E$ associated with the entity in $t=1$, and
3) whether the entity has been deplatformed ($D$). 
Our methodology is unable to distinguish between the causal effects associated with $E\rightarrow A_{t=2}$ (which we refer to as $\alpha$) and $D\rightarrow A_{t=2}$ (which we refer to as $\beta$), estimating both jointly ($\alpha + \beta$). Yet, as the total effect is negative ($\alpha + \beta < 0$) and as the effect from the event is likely to increase online attention ($\alpha > 0$), it follows that the real effect $\beta$ of deplatforming must be negative, and smaller or equal to the estimated effect $\alpha + \beta$.

\xhdr{Beyond engagement on social media platforms}
An important way our research differs from previous work is that we go beyond social-media indicators of attention: we examine changes in overall online attention to these entities through search engine behavior and Wikipedia pageviews. 
This is, in part, what allowed us to circumvent the methodological limitations of previous work~\cite{russo2023spillover, horta2021platform, jhaver2021evaluating, mitts2021banned, trujillo2022make, trujillo_one_2023}. 
Likes, posts, and comments are platform-specific and may underestimate the popularity of fringe influencers that migrate to alternative platforms (e.g., Gab, Rumble). 
In contrast, online attention traces like the ones we used are arguably better suited to capture the impact of moderation decisions on the broader information ``ecosystem.''

Researching social media has become increasingly difficult in what Freelon has named ``the post-API age,'' the systematic degradation of infrastructure researchers use to study online platforms~\cite{freelon2018computational}.
Lazer~\cite{lazer2020studying} wrote that 
``{the Internet should be viewed as an [\dots] experiment, manipulating what people see and how they see it. The access that science has to this information is quite limited, however}.''
Here, we consider holistic attention across the Web while studying phenomena inherently linked to social media. 
Other researchers, too, may use this approach to ``dip their toes'' in online attention data without partnering with social media companies.

Another approach that has been growing in popularity is the use of online panels captured by media companies~\cite{horta2023deplatforming,agarwal2022deplatforming}, where panelist, paid or voluntary, install browser extensions or mobile phone apps that allow companies and researchers to peek into their online habits, \eg, analyzing their web history.
In contrast to this strategy, using Google Trends and Wikipedia Pageviews is easier and remarkably cheaper.
Nonetheless, it is worth noting the goal of initiatives such as the Observatory for Online Human and Platform Behavior~\cite{northeastern_observatory_2023} is to provide researchers with access to this data for free, so this can be a convenient way to conduct similar research in future work (the authors of this paper are not involved in this initiative).

\xhdr{Deliberate \vs coincidental attention}
Our findings are in contrast with, but do not contradict, previous work analyzing the impact of deplatforming Parler, a far-right social media platform, from Amazon Web Hosting Services~\cite{horta2023deplatforming,agarwal2022deplatforming}.
Previous work has found that banning Parler drove users to other alternative social media platforms (\eg, Gab, Bitchute), such that the overall activity across platforms remained the same~\cite{horta2023deplatforming,agarwal2022deplatforming}.
At the same time, past research has found that engagement with radical content on YouTube is driven primarily by self-selection~\cite{hosseinmardi2023causally,hosseinmardi2021examining, chen2023subscriptions}.
These findings suggest a fundamental difference between deplatforming influencers exposed to everyday users that do not necessarily wish to engage with them on mainstream platforms but \textit{coincidentally} do so (\eg, Alex Jones on Twitter) and influencers who are engaging mostly with a dedicated audience, that \textit{deliberately} go out of their way to consume their content (\eg, Alex Jones on Parler). 
Considering these previous findings, we conjecture that deplatforming is most effective in decreasing ``spontaneous'' attention, but further exploring these different ``kinds'' of online attention might be an interesting avenue for future work.

\xhdr{Continuously tracking moderation interventions}
Social media platforms such as Facebook, YouTube, Twitter, and Reddit have changed the fabric of society. In part, the effects of these platforms on society are mediated by how they moderate content. 
Thus, we argue that comprehensive research examining the impact of moderation practices (\eg, deplatforming) is key so that platform decisions are not made in a vacuum or solely to preserve the platforms' own financial interests.
An essential element of this research agenda is the continued monitoring of the effectiveness of moderation interventions. Future work could continue to track and analyze moderation sanctions using the methodology provided here.
It would be particularly interesting to see if there are differences across countries and across years as the digital ecosystem evolves.

\xhdr{Ethical considerations} In this work, we only used data publicly available on the Web and did not interact with online users in any way. We study and make several deplatforming events available. Yet, each event was covered in news pieces and widely discussed on Reddit, leading us to believe that we are not infringing on reasonable privacy expectations.









\bibliographystyle{acm}
\bibliography{main}

\appendix
\begin{table}[]
\centering
\caption{Final patterns used to extract deplatforming events.}
\label{tab:final_patterns}
\begin{tabular}{l}
\toprule
\textbf{Patterns}\\ \midrule
<ent> $\times$ on <plat> \\ \midrule
<plat> employee deactivated <ent>\\ \midrule
<plat> employee `deactivated` <ent>\\  \midrule
<plat> employee deactivates <ent>\\  \midrule
<plat> bans <ent>\\  \midrule
<plat> suspends <ent>\\  \midrule
<plat> suspended <ent>\\  \midrule
<plat> banned <ent>	\\ \midrule
<plat> removes <ent>\\	 \midrule
<plat> ban <ent>\\ \midrule
<plat> banning <ent>\\ \midrule
<ent> locked out of <plat>	\\ \midrule
<ent> blocked on <plat>\\ \midrule
<plat> deleted <ent>\\ \midrule
<plat> takes down <ent>	\\ \midrule
<plat> blocks <ent>	\\ \bottomrule
\end{tabular}

\end{table}
\end{document}
\endinput

%% file: dlabmacros.tex
\usepackage[utf8]{inputenc}
\usepackage[T1]{fontenc}
\usepackage{hyphenat}
\usepackage{xspace}
\usepackage{amsmath}
\usepackage{amsfonts}
\usepackage{hyperref}
\usepackage{url}
\usepackage{booktabs}
\usepackage{multirow}
\usepackage{makecell}
\usepackage{caption}
\usepackage{minibox}
\usepackage{bbm}
\usepackage{graphicx}
\usepackage{balance}
\usepackage{mathtools}
\usepackage{color}
\usepackage{marvosym}
\usepackage{ifthen}
\usepackage{textcomp}
\usepackage{enumitem}
\usepackage{verbatim}
\usepackage{algorithm}
\usepackage{algorithmic}
\usepackage{numprint}
\usepackage{balance}

\usepackage{amsthm}
\theoremstyle{plain}

\newcommand{\chatoDisplayMode}[1]{#1}

\definecolor{MyRed}{rgb}{0.6,0.0,0.0} 
\definecolor{MyBlack}{rgb}{0.1,0.1,0.1} 
\newcommand{\inred}[1]{{\color{MyRed}\sf\textbf{\textsc{#1}}}}
\newcommand{\frameit}[2]{
  \begin{center}
  {\color{MyRed}
  \framebox[.9\columnwidth][l]{
    \begin{minipage}{.85\columnwidth}
    \inred{#1}: {\sf\color{MyBlack}#2}
    \end{minipage}
  }\\
  }
  \end{center}
}

\newcommand{\note}[2][]{\chatoDisplayMode{\def\@tmpsig{#1}\frameit{{\Pointinghand} Note}{#2\ifx \@tmpsig \@empty \else \mbox{ --\em #1}\fi}}}
\newcommand{\todo}[2][]{\chatoDisplayMode{\def\@tmpsig{#1}\frameit{{\Writinghand} To-do}{#2\ifx \@tmpsig \@empty \else \mbox{ --\em #1}\fi}}}

\newcommand{\abbrevStyle}[1]{#1}

\newcommand{\ie}{\abbrevStyle{i.e.}\xspace}
\newcommand{\eg}{\abbrevStyle{e.g.}\xspace}
\newcommand{\cf}{\abbrevStyle{cf.}\xspace}

\newcommand{\vs}{\abbrevStyle{vs.}\xspace}
\newcommand{\etc}{\abbrevStyle{etc.}\xspace}

\newcommand{\Secref}[1]{Sec.~\ref{#1}}

\newcommand{\Tabref}[1]{Table~\ref{#1}}
\newcommand{\Figref}[1]{Fig.~\ref{#1}}

\newcommand{\xhdr}[1]{\vspace{1.7mm}\noindent{{\bf #1.}}}

\newcommand{\xhdrNoPeriod}[1]{\vspace{1.7mm}\noindent{{\bf #1}}}

\newcommand{\textcite}[1]{\citeauthor{#1} \shortcite{#1}}

\newcommand{\hide}[1]{}

\makeatletter
\newcommand{\iffont}[2]{\ifthenelse{\equal{\f@family}{#1}}{#2}{}}
\makeatother

\iffont{ptm}{
  \usepackage{mathptmx}

  \DeclareSymbolFont{greek}{OML}{cmm}{m}{n}
  \DeclareMathSymbol{\alpha}{\mathalpha}{greek}{"0B}
  \DeclareMathSymbol{\beta}{\mathalpha}{greek}{"0C}
  \DeclareMathSymbol{\gamma}{\mathalpha}{greek}{"0D}
  \DeclareMathSymbol{\delta}{\mathalpha}{greek}{"0E}
  \DeclareMathSymbol{\epsilon}{\mathalpha}{greek}{"0F}
  \DeclareMathSymbol{\zeta}{\mathalpha}{greek}{"10}
  \DeclareMathSymbol{\eta}{\mathalpha}{greek}{"11}
  \DeclareMathSymbol{\theta}{\mathalpha}{greek}{"12}
  \DeclareMathSymbol{\iota}{\mathalpha}{greek}{"13}
  \DeclareMathSymbol{\kappa}{\mathalpha}{greek}{"14}
  \DeclareMathSymbol{\lambda}{\mathalpha}{greek}{"15}
  \DeclareMathSymbol{\mu}{\mathalpha}{greek}{"16}
  \DeclareMathSymbol{\nu}{\mathalpha}{greek}{"17}
  \DeclareMathSymbol{\xi}{\mathalpha}{greek}{"18}
  \DeclareMathSymbol{\pi}{\mathalpha}{greek}{"19}
  \DeclareMathSymbol{\rho}{\mathalpha}{greek}{"1A}
  \DeclareMathSymbol{\sigma}{\mathalpha}{greek}{"1B}
  \DeclareMathSymbol{\tau}{\mathalpha}{greek}{"1C}
  \DeclareMathSymbol{\upsilon}{\mathalpha}{greek}{"1D}
  \DeclareMathSymbol{\phi}{\mathalpha}{greek}{"1E}
  \DeclareMathSymbol{\chi}{\mathalpha}{greek}{"1F}
  \DeclareMathSymbol{\psi}{\mathalpha}{greek}{"20}
  \DeclareMathSymbol{\omega}{\mathalpha}{greek}{"21}
  \DeclareMathSymbol{\varepsilon}{\mathalpha}{greek}{"22}
  \DeclareMathSymbol{\vartheta}{\mathalpha}{greek}{"23}
  \DeclareMathSymbol{\varpi}{\mathalpha}{greek}{"24}
  \DeclareMathSymbol{\varrho}{\mathalpha}{greek}{"25}
  \DeclareMathSymbol{\varsigma}{\mathalpha}{greek}{"26}
  \DeclareMathSymbol{\varphi}{\mathalpha}{greek}{"27}
  \DeclareSymbolFont{otone}{OT1}{cmr}{m}{n}
  \DeclareMathSymbol{\Gamma}{\mathalpha}{otone}{0}
  \DeclareMathSymbol{\Delta}{\mathalpha}{otone}{1}
  \DeclareMathSymbol{\Theta}{\mathalpha}{otone}{2}
  \DeclareMathSymbol{\Lambda}{\mathalpha}{otone}{3}
  \DeclareMathSymbol{\Xi}{\mathalpha}{otone}{4}
  \DeclareMathSymbol{\Pi}{\mathalpha}{otone}{5}
  \DeclareMathSymbol{\Sigma}{\mathalpha}{otone}{6}
  \DeclareMathSymbol{\Upsilon}{\mathalpha}{otone}{7}
  \DeclareMathSymbol{\Phi}{\mathalpha}{otone}{8}
  \DeclareMathSymbol{\Psi}{\mathalpha}{otone}{9}
  \DeclareMathSymbol{\Omega}{\mathalpha}{otone}{10}
  \DeclareSymbolFont{syms}{OML}{cmm}{m}{it}
  \DeclareMathSymbol{\partial}{\mathord}{syms}{"40}
  \DeclareMathAlphabet{\mathbold}{OML}{cmm}{b}{it}
  \DeclareSymbolFont{largesymbols}{OMX}{cmex}{m}{n}

}

%% file: tableeff.tex
    \setlength\tabcolsep{10pt}
    \setlength\extrarowheight{2pt} 
\begin{tabular}{lllll}
\toprule
 &  &  & Effect & $p$-value \\
Model & Variable & Source &  &  \\
\midrule
\multirow[t]{2}{*}{1} & \multirow[t]{2}{*}{Temporary} & G-Trends & 13.2\% 95\% CI [-45.5\%, 135.1\%] & 0.739 \\
 &  & Wikipedia & -5.3\% 95\% CI [-36.5\%, 41.3\%] & 0.791 \\
\midrule
\multirow[t]{2}{*}{2} & \multirow[t]{2}{*}{High attention} & G-Trends & -58.8\% 95\% CI [-72.6\%, -37.9\%] & $<$0.001 \\
 &  & Wikipedia & -33.7\% 95\% CI [-48.2\%, -15.1\%] & 0.001 \\
\midrule
\multirow[t]{4}{*}{3} & \multirow[t]{2}{*}{Hate} & G-Trends & -19.5\% 95\% CI [-57.2\%, 51.4\%] & 0.501 \\
 &  & Wikipedia & -20.8\% 95\% CI [-46.6\%, 17.5\%] & 0.247 \\
\cline{2-5}
 & \multirow[t]{2}{*}{Manipulation} & G-Trends & -59.8\% 95\% CI [-82.4\%, -7.7\%] & 0.032 \\
 &  & Wikipedia & -48.7\% 95\% CI [-67.5\%, -19.2\%] & 0.004 \\
\midrule
\multirow[t]{8}{*}{4} & \multirow[t]{2}{*}{Politician} & G-Trends & -5.5\% 95\% CI [-63.5\%, 144.7\%] & 0.907 \\
 &  & Wikipedia & -0.9\% 95\% CI [-44.5\%, 76.7\%] & 0.975 \\
\cline{2-5}
 & \multirow[t]{2}{*}{Media pers.} & G-Trends & 32.2\% 95\% CI [-44.7\%, 215.8\%] & 0.530 \\
 &  & Wikipedia & 22.8\% 95\% CI [-24.1\%, 98.5\%] & 0.403 \\
\cline{2-5}
 & \multirow[t]{2}{*}{Internet pers.} & G-Trends & 25.1\% 95\% CI [-45.1\%, 184.8\%] & 0.594 \\
 &  & Wikipedia & 35.6\% 95\% CI [-24.4\%, 143.2\%] & 0.308 \\
\cline{2-5}
 & \multirow[t]{2}{*}{Fringe mov.} & G-Trends & -38.6\% 95\% CI [-71.8\%, 33.8\%] & 0.220 \\
 &  & Wikipedia & -1.4\% 95\% CI [-34.0\%, 47.2\%] & 0.945 \\
\bottomrule
\end{tabular}